\documentclass[twocolumn]{aastex62}
\usepackage{CJK}
\usepackage{graphicx}

\shorttitle{Partial Eruption of Solar Filaments}
\shortauthors{Hou et al.}

\begin{document}

\title{Partial Eruption of Solar Filaments. I. Configuration and Formation of Double-decker Filaments}

\correspondingauthor{Yijun Hou}
\email{yijunhou@nao.cas.cn}

\author[0000-0002-9534-1638]{Yijun Hou}
\affiliation{National Astronomical Observatories, Chinese Academy of Sciences, Beijing 100101, China}
\affiliation{Yunnan Key Laboratory of the Solar physics and Space Science, Kunming, 650216, China}
\affiliation{State Key Laboratory of Solar Activity and Space Weather, Beijing, 100190, China}
\affiliation{School of Astronomy and Space Science, University of Chinese Academy of Sciences, Beijing 100049, China}

\author[0000-0001-7693-4908]{Chuan Li}
\affiliation{School of Astronomy and Space Science, Nanjing University, Nanjing 210023, China}
\affiliation{Key Laboratory of Modern Astronomy and Astrophysics (Nanjing University), Ministry of Education, Nanjing 210023, China}

\author[0000-0001-6655-1743]{Ting Li}
\affiliation{National Astronomical Observatories, Chinese Academy of Sciences, Beijing 100101, China}
\affiliation{State Key Laboratory of Solar Activity and Space Weather, Beijing, 100190, China}
\affiliation{School of Astronomy and Space Science, University of Chinese Academy of Sciences, Beijing 100049, China}

\author{Jiangtao Su}
\affiliation{National Astronomical Observatories, Chinese Academy of Sciences, Beijing 100101, China}
\affiliation{State Key Laboratory of Solar Activity and Space Weather, Beijing, 100190, China}
\affiliation{School of Astronomy and Space Science, University of Chinese Academy of Sciences, Beijing 100049, China}

\author{Ye Qiu}
\affiliation{School of Astronomy and Space Science, Nanjing University, Nanjing 210023, China}
\affiliation{Key Laboratory of Modern Astronomy and Astrophysics (Nanjing University), Ministry of Education, Nanjing 210023, China}

\author[0000-0002-6565-3251]{Shuhong Yang}
\affiliation{National Astronomical Observatories, Chinese Academy of Sciences, Beijing 100101, China}
\affiliation{State Key Laboratory of Solar Activity and Space Weather, Beijing, 100190, China}
\affiliation{School of Astronomy and Space Science, University of Chinese Academy of Sciences, Beijing 100049, China}

\author{Liheng Yang}
\affiliation{Yunnan Observatories, Chinese Academy of Sciences, Kunming, 650216, China}
\affiliation{Yunnan Key Laboratory of the Solar physics and Space Science, Kunming, 650216, China}

\author{Leping Li}
\affiliation{National Astronomical Observatories, Chinese Academy of Sciences, Beijing 100101, China}
\affiliation{State Key Laboratory of Solar Activity and Space Weather, Beijing, 100190, China}
\affiliation{School of Astronomy and Space Science, University of Chinese Academy of Sciences, Beijing 100049, China}

\author[0000-0001-9893-1281]{Yilin Guo}
\affiliation{Beijing Planetarium, Beijing Academy of Science and Technology, Beijing 100044, China}

\author{Zhengyong Hou}
\affiliation{School of Earth and Space Sciences, Peking University, Beijing 100871, China}

\author{Qiao Song}
\affiliation{Key Laboratory of Space Weather, National Satellite Meteorological Center (National Center for Space Weather),
China Meteorological Administration, Beijing 100081, China}
\affiliation{Innovation Center for FengYun Meteorological Satellite (FYSIC), Beijing 100081, China}

\author{Xianyong Bai}
\affiliation{National Astronomical Observatories, Chinese Academy of Sciences, Beijing 100101, China}
\affiliation{State Key Laboratory of Solar Activity and Space Weather, Beijing, 100190, China}
\affiliation{School of Astronomy and Space Science, University of Chinese Academy of Sciences, Beijing 100049, China}

\author{Guiping Zhou}
\affiliation{National Astronomical Observatories, Chinese Academy of Sciences, Beijing 100101, China}
\affiliation{State Key Laboratory of Solar Activity and Space Weather, Beijing, 100190, China}
\affiliation{School of Astronomy and Space Science, University of Chinese Academy of Sciences, Beijing 100049, China}

\author{Mingde Ding}
\affiliation{School of Astronomy and Space Science, Nanjing University, Nanjing 210023, China}
\affiliation{Key Laboratory of Modern Astronomy and Astrophysics (Nanjing University), Ministry of Education, Nanjing 210023, China}

\author{Weiqun Gan}
\affiliation{Purple Mountain Observatory, Chinese Academy of Sciences, Nanjing 210034, China}

\author{Yuanyong Deng}
\affiliation{National Astronomical Observatories, Chinese Academy of Sciences, Beijing 100101, China}
\affiliation{State Key Laboratory of Solar Activity and Space Weather, Beijing, 100190, China}
\affiliation{School of Astronomy and Space Science, University of Chinese Academy of Sciences, Beijing 100049, China}

\begin{abstract}
Partial eruptions of solar filaments are the typical representative of solar eruptive behavior diversity. Here we investigate a
typical filament partial eruption event and present integrated evidence for configuration of the pre-eruption filament and its
formation. The \emph{CHASE} H$\alpha$ observations reveal structured Doppler velocity distribution within the pre-eruption filament,
where distinct redshift only appeared in the east narrow part of the south filament region and then disappeared after the partial
eruption while the north part dominated by blueshift remained. Combining the \emph{SDO}, \emph{ASO-S} observations, and NLFFF
modeling results, we verify that there were two independent material flow systems within the pre-flare filament, whose magnetic
topology is a special double-decker configuration consisting of two magnetic flux ropes (MFRs) with opposite magnetic twist.
During the formation of this filament system, continuous magnetic flux cancellation and footpoint motion were observed around
its north end. Therefore, we propose a new double-decker formation scenario that the two MFRs composing such double-decker
configuration originated from two magnetic systems with different initial connections and opposite magnetic twist. Subsequent
magnetic reconnection with surrounding newly-emerging fields resulted in the motion of footpoint of the upper MFR to the region
around footpoint of the lower MFR, thus leading to eventual formation of the double-decker configuration consisting of two MFRs
with similar footpoints but opposite signs of magnetic twist. These results provide a potential way to determine unambiguously
the progenitor configuration of a partial-eruptive filament and reveal a special type of double-decker MFR configuration and
a new double-decker formation scenario.
\end{abstract}


\section{Introduction}\label{sect1}
On the solar disk, elongated dark absorption features are often seen in H$\alpha$ and extreme ultraviolet (EUV) observations
and are called ``solar filaments". When moving to the solar limb, solar filaments appear as bright emission structures suspended
in the corona and are described as ``prominences" \citep{2010SSRv..151..333M,2020RAA....20..166C}. Solar filaments are related
to abundant physical processes and multi-scale activities occurring in solar atmosphere, e.g., thermal instabilities during the
filament formation \citep{1991ApJ...378..372A,2011ApJ...737...27X,2020NatAs...4..994Z}, oscillations of the filament threads
\citep{2002SoPh..206...45O,2012ApJ...760L..10L,2015ApJ...809...72A}, and various types of small-scale activities induced by
magnetohydrodynamic (MHD) instabilities or magnetic reconnection \citep{2015ApJ...814L..17S,2017ApJ...850...60B,
2018ApJ...863..192L,2022A&A...659A..76W,2022NatAs...6..942J}. Especially, filament eruptions are closely related to solar
flares and coronal mass ejections (CMEs), which can pose disastrous disturbances to the space environment near the Earth.
As a result, solar filaments are the ideal object for studying magnetic and plasma structures of corona and their evolutions,
mass and energy transportations between different solar atmospheric layers, as well as the driving mechanisms of solar eruptions.

Solar filaments usually lie along the polarity inversion line (PIL) separating photospheric magnetic fields with opposite
radial components. As regards the magnetic topology of solar filaments, there are two typical models: the sheared arcade
model for the normal-polarity filaments and the magnetic flux rope (MFR, a helical structure of magnetic field lines wrapping
more than once around a central axis) model for the inverse-polarity filaments \citep{2002ApJ...567L..97A,2010SSRv..151..333M,
2016ApJ...818..148L}. In both models, the filament plasmas are pilled at magnetic dips, where magnetic tension supports plasmas
against their gravity. It is worth noting that an MFR and a sheared arcade could co-exist in one solar filament and match two
sections of the filament, respectively \citep{2010ApJ...714..343G}.

Triggered by several possible mechanisms, e.g., magnetic-reconnection-related processes \citep{2000ApJ...545..524C,
2001ApJ...548L..99Z,2004ApJ...602.1024S,2018ApJ...862..117D,2020ApJ...892...79S,2020A&A...640A.101H} and ideal MHD instabilities
\citep{2005ApJ...630L..97T,2010ApJ...718.1388D,2017ApJ...849L..21Y,2018A&A...619A.100H,2020ApJ...890...10Z}, filaments could
lose the balance of forces acting on them and eventually erupt. In the classical solar eruption scenarios (e.g., CSHKP model),
filament eruptions play a critical role in the onset of solar flares and CMEs. It is well established that an erupting filament
will push its overlying magnetic fields upwards and form a reconnection region below, where particles are accelerated and
propagate downward along the newly formed magnetic fields, hitting the lower solar atmosphere and producing flare ribbons.
When the erupting filament successfully escapes the solar atmosphere, a CME would be produced \citep{2000JGR...105.2375L,
2002A&ARv..10..313P,2013AdSpR..51.1967S}.

Previous high-resolution observations have revealed that solar filaments exhibit a wide range of eruptive dynamics.
\citet{2007SoPh..245..287G} developed observational definitions for three types of filament eruptions: (1) ``full eruption", the
magnetic structure and material of pre-eruptive filament completely escape the Sun; (2) ``failed eruption", the eruptive process
of filament is suddenly halted in the low corona, with none of the lifted filament material nor magnetic structure escaping the
Sun \citep{2003ApJ...595L.135J,2005ApJ...630L..97T,2015Natur.528..526M,2017ApJ...838...15L,2020ApJ...889..106Y}; 
(3) ``partial eruption", only part of the filament magnetic structure and/or material is eventually expelled 
\citep{2000ApJ...537..503G,2015ApJ...805....4Z,2015ApJ...805...48B,2018ApJ...869...78C,2022ApJ...929...85D}. Among the three types, 
filament partial eruptions are relatively more complicated because the involved pre-eruption filament does not evolve as a whole 
and shows nonuniform characteristics in terms of magnetic topology and/or plasma motion.

Regarding the mechanism of filament partial eruptions, two main physical scenarios are proposed by observational and numerical studies.
The first one was proposed by \citet{2006ApJ...637L..65G} through three-dimensional (3D) MHD modeling that the magnetic structure
of the pre-eruption filament is an MFR with a bald patch (BP), where the bottom magnetic field lines of MFR touch the photosphere
tangentially. During the eruption of this MFR, the magnetic field lines tying in the BP prevent the lower part of the MFR from erupting,
naturally leading to internal magnetic reconnection within the MFR and subsequent vertical splitting of the MFR into two parts, with
one part successfully being expelled and the other one remaining behind \citep{2001ApJ...549.1221G}. This mechanism has been further
supported by observations of splitting process accompanied by obvious local brightening between the erupting and remaining parts of
the filament \citep{2009A&A...498..295T,2013ApJ...778..142T,2018ApJ...856...48C,2022ApJ...940L..12H,2023ApJ...953..148S}. In addition,
solar filaments are also observed to split under the interaction with surrounding magnetic structures. For example,
\citet{2023ApJ...945....5X} reported a filament splitting event caused by the reconnection of emerging magnetic fields with one leg of
the pre-eruption filament. Similar splitting processes driven by such reconnection with surrounding structures could also further result
in partial eruptions of solar filaments \citep{2018ApJ...869...78C,2021MNRAS.500..684M,2022ApJ...929...85D}.

The second scenario for the filament partial eruption is the double-decker model proposed by \citet{2012ApJ...756...59L}, which was
later simulated by \citet{2014ApJ...792..107K}. In this model, the magnetic structure of the pre-eruption filament consists of two
vertically-distributed parts: two separate MFRs or one MFR above sheared arcades. Observations have revealed that such partially
erupting filament is composed of two branches separated in height, which have already existed several hours prior to the eruption.
Intermittent transfers of magnetic flux and current from the lower branch to the upper one serve as the key mechanism for the upper
branch to lose equilibrium and erupt eventually \citep{2012ApJ...756...59L,2014SoPh..289..279Z}. This scenario has also been supported
by observations and nonlinear-force-free-field (NLFFF) extrapolation results in following studies \citep{2014ApJ...789...93C,
2018A&A...619A.100H,2020IAUS..354..443H,2019ApJ...872..109A,2019ApJ...875...71Z,2020ApJ...900...23M,2021ApJ...909...32P,2021ApJ...923..142C,2022ApJ...933..200Z}.

Although the partial eruptions of solar filaments have been extensively investigated over the past decades, our knowledge of this kind
of filament eruption is continuously updated, and several open questions still remain: How can we determine unambiguously which mechanism
is responsible for the observed filament partial eruption? Can the two mechanisms introduced above simultaneously play roles during the
partial eruption of filament? Is there another type of double-decker configuration different from the typical ones existing in the
filament partial eruption event? How does this type of double-decker configuration form and evolve into eruption? Insights into these
questions are necessary to further understand the eruptive behavior diversity and magnetic topology complexity of solar filaments.

In the present work, we investigate a typical filament partial eruption event and present integrated evidence for the double-decker
configuration of pre-eruption filament and its formation process from the aspects of material flow pattern, magnetic topology, and
their morphological evolutions. The data from the \emph{Chinese H$\alpha$ Solar Explorer} \citep[\emph{CHASE};][]{2022SCPMA..6589602L},
\emph{Solar Dynamics Observatory} \citep[\emph{SDO};][]{2012SoPh..275....3P}, \emph{Advanced Space-based Solar Observatory}
\citep[\emph{ASO-S};][]{2023SoPh..298...68G}, and other instruments are analyzed, aided by the NLFFF extrapolations with a time sequence,
to reveal configuration and formation of this double-decker filament system. The remainder of this paper is structured as follows.
Section \ref{sect2} describes the observations and data analysis methods taken in our study. In Section \ref{sect3}, the results of
observations and analysis are presented and discussed. Finally, we briefly summarize the major findings in Section \ref{sect4}.

\begin{table*}
\caption{Details of the data analyzed in this study\label{t1}}
\centering
\begin{tabular}{c  c  c  c  c}   
\hline\hline
\textbf{Telescope} & \textbf{Time (UT)} & \textbf{Passband \& Data product} & \textbf{Cadence} & \textbf{Spatial \& Spectral resolutions} \\
\hline
\emph{CHASE}/HIS & 18:48--19:04 & Imaging data of H$\alpha$ 6562.8 {\AA} & $\sim$71 s & $\sim$1.0{\arcsec} pixel$^{-1}$ \\
                 &              & Spectra data of H$\alpha$ 6562.8 {\AA} & $\sim$71 s & $\sim$0.048 {\AA} pixel$^{-1}$  \\
                 & 20:23--20:39 & Imaging data of H$\alpha$ 6562.8 {\AA} & $\sim$71 s & $\sim$1.0{\arcsec} pixel$^{-1}$ \\
                 &              & Spectra data of H$\alpha$ 6562.8 {\AA} & $\sim$71 s & $\sim$0.048 {\AA} pixel$^{-1}$  \\
\hline
\emph{ASO-S}/FMG & 10:07--20:30 & LOS magnetogram & $\sim$116 s & $\sim$0.5{\arcsec} pixel$^{-1}$  \\
                 & 10:07--20:30 & Filtergram of Fe {\sc i} 5324.2 {\AA}  & $\sim$116 s & $\sim$0.5{\arcsec} pixel$^{-1}$ \\
\hline
\emph{SDO}/AIA   & 00:00--20:30 & 304, 171, 94 {\AA} & 12 s & $\sim$0.6{\arcsec} pixel$^{-1}$ \\
\hline
\emph{SDO}/HMI   & 00:00--20:30 & Full-disk vector magnetograms & 720 s & $\sim$0.5{\arcsec} pixel$^{-1}$ \\
                 & 00:00--20:30 & LOS magnetogram & 45 s & $\sim$0.5{\arcsec} pixel$^{-1}$ \\
\hline
\emph{SATech-01}/SUTRI & 19:38 & 465 {\AA} & / & $\sim$1.2{\arcsec} pixel$^{-1}$ \\
\hline
\emph{FY-3E}/X-EUVI      & 19:43 & 195 {\AA} & / & $\sim$2.4{\arcsec} pixel$^{-1}$ \\
\hline
\emph{GOES}/EXIS       & 18:30--21:00 &  soft X-ray (SXR) flux of 1--8 {\AA} & 60 s & / \\
\hline
\end{tabular}
\end{table*}

\section{Observations and data analysis}\label{sect2}
The filament partial eruption event of interest occurred in NOAA AR 13176 on 2022 December 30 and was well observed by the
\emph{CHASE}, \emph{SDO}, and \emph{ASO-S}. Additionally, the EUV images taken by the Solar Upper Transition Region Imager
\citep[SUTRI;][]{2023RAA....23f5014B} on board the first spacecraft of the Space Advanced Technology demonstration satellite
series (\emph{SATech-01}), Chinese Academy of Sciences (CAS) and the Solar X-ray and Extreme Ultraviolet Imager (\emph{X-EUVI})
on board the \emph{FengYun-3E} \citep[\emph{FY-3E};][]{2022AdAtS..39....1Z}, satellite, and the soft X-ray (SXR) 1--8 {\AA}
flux from the \emph{GOES} were also employed. The details of these observations are summarized in Table \ref{t1}.

The \emph{CHASE}, launched on 2021 October 14, is the first solar space mission of China National Space Administration (CNSA). The
H$\alpha$ Imaging Spectrograph \citep[HIS;][]{2022SCPMA..6589605L} is the scientific payload of the \emph{CHASE} and can provide
spectroscopic observations of the Sun by scanning the full solar disk in both H$\alpha$ (6559.7--6565.9 {\AA}) and Fe {\sc i}
(6567.8--6570.6 {\AA}) wavebands. The \emph{CHASE}/HIS data applied here have been calibrated through dark-field, flat-field,
and slit image curvature corrections, wavelength and intensity calibration, and coordinate transformation \citep{2022SCPMA..6589603Q}.
Since the H$\alpha$ line profiles are roughly symmetrical, we adopted the moment analysis to obtain Doppler velocity of the filament,
i.e., $\lambda_{0}=\frac{\sum\lambda_{i} (I_{i}-I_{c})}{\Sigma (I_{i}-I_{c})}$, where $I_c$ is the continuum intensity, $I_{i}$ and
$\lambda_{0}$ are the intensity and calculated gravity center of the observed H$\alpha$ line profile. The zero Doppler shift is
defined as the gravity center of the averaged profile in a quiescent region [-80{\arcsec}, -68{\arcsec}]$\times$[385{\arcsec}, 395{\arcsec}].
The uncertainty is then estimated as the root mean square of velocity of this region, which are 0.37 $\rm km~s^{-1}$ and
0.36 $\rm km~s^{-1}$ at 19:03:07 UT and 20:27:24 UT, respectively.

The \emph{ASO-S} is the first comprehensive Chinese dedicated solar observatory in space and its primary scientific objective is
to improve our understanding of solar flares, CMEs, solar magnetic fields, and the relationships among them. The Full-disk vector
MagnetoGraph \citep[FMG;][]{2019RAA....19..157D} on board the \emph{ASO-S} is designed to measure the solar photospheric magnetic
fields through Fe {\sc i} 532.42 nm line with high spatial and temporal resolutions, and high magnetic sensitivity. Here we utilized
the line-of-sight (LOS) magnetograms and filtergram of Fe {\sc i} 532.42 nm line taken by the \emph{ASO-S}/FMG.
Furthermore, the EUV images at Ne {\sc vii} 46.5 nm with a formation temperature of $\sim$0.5 MK and Fe {\sc xii} 19.5 nm line of
$\sim$1.5 MK observed by \emph{SATech-01}/SUTRI and \emph{FY-3E}/X-EUVI, respectively, are also analyzed to show the M3.7 flare
caused by the filament partial eruption event.

The Atmospheric Imaging Assembly \citep[AIA;][]{2012SoPh..275...17L} and Helioseismic and Magnetic Imager \citep[HMI;][]{2012SoPh..275..229S}
on board the \emph{SDO} can successively observe the multilayered solar atmosphere in ten passbands and the photospheric magnetic field,
respectively. Here, we analyzed the AIA 304 {\AA}, 171 {\AA}, and 94 {\AA} images as well as the HMI LOS magnetograms and photospheric
vector magnetic field. The \emph{CHASE}, \emph{SDO}, \emph{ASO-S}, SUTRI, and X-EUVI observations are carefully co-aligned by matching
locations of some specific features that can be simultaneously detected in different channels of these telescopes.

In order to reconstruct the 3D magnetic configuration above the region of interest, we utilized the ``weighted optimization" method to
perform NLFFF extrapolation \citep{2004SoPh..219...87W,2012SoPh..281...37W} based on the photospheric vector magnetic fields observed by
\emph{SDO}/HMI. Before extrapolation calculation, the HMI vector magnetograms were preprocessed by a procedure developed by
\citet{2006SoPh..233..215W} towards suitable photospheric boundary conditions consistent with the force-free assumption. Then the
calculations were performed within a box of 720$\times$448$\times$256 uniform grid points (261$\times$163$\times$93 Mm$^{3}$),
which covers nearly the entire AR. Furthermore, through the method developed by \citet{2016ApJ...818..148L}, we calculated the twist
number $T_{w}$ and squashing factor $Q$ of the extrapolated 3D magnetic fields. The squashing factor $Q$ provides important information
about the magnetic connectivity, and the twist number $T_{w}$ represents how many turns two field lines wind about each other and
plays a critical role in identifying an MFR without ambiguity.

\section{Results and discussion}\label{sect3}
\subsection{Overview of the filament partial eruption event}\label{sect31}
\begin{figure*}
\centering
\includegraphics [width=0.99\textwidth]{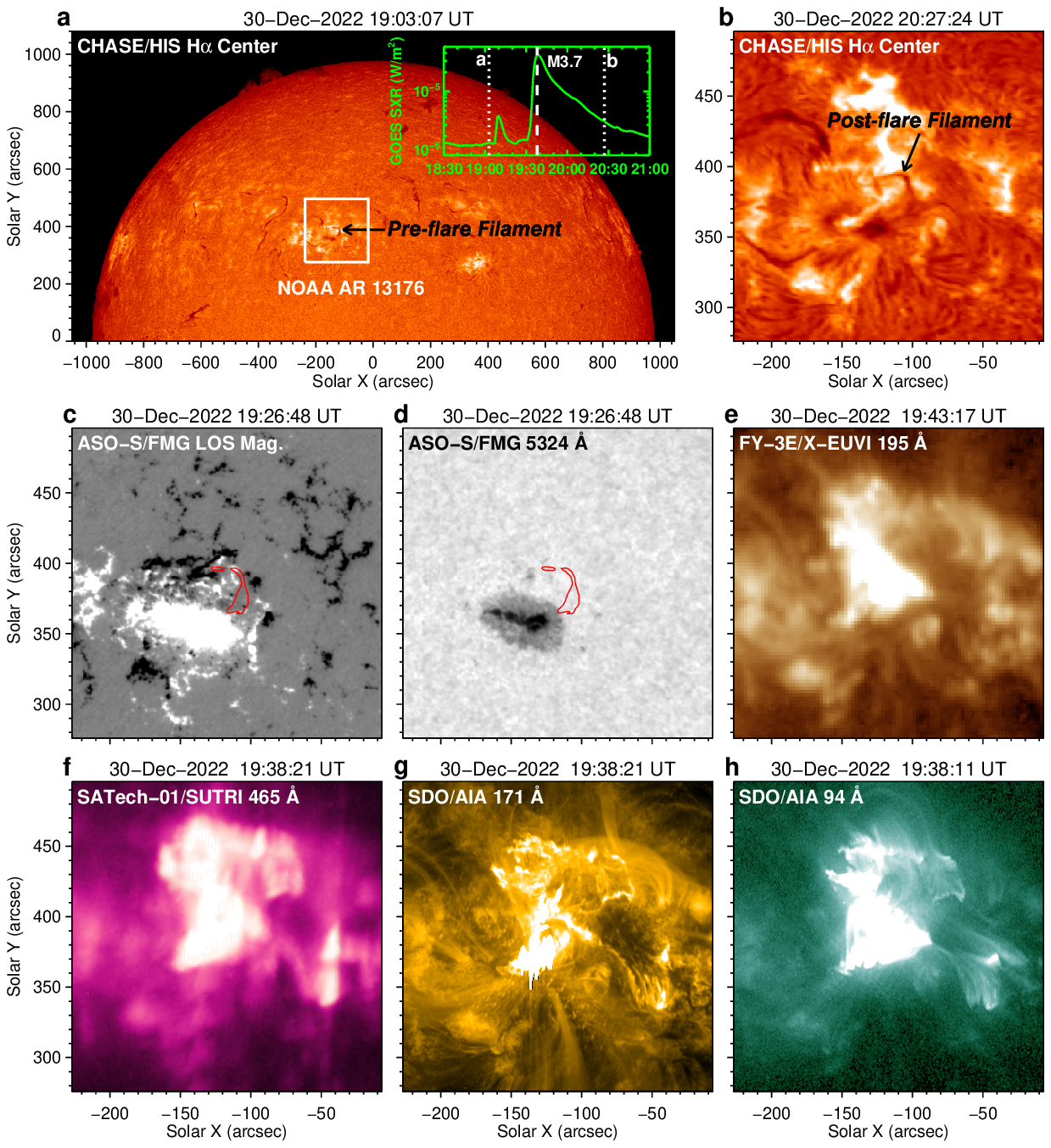}
\caption{Overview of the filament partial eruption event and the resultant M3.7 flare on 2022 December 30.
(a)--(b): \emph{CHASE}/HIS H$\alpha$ center (6562.8 {\AA}) images showing the filament of interest before and after the M3.7 flare.
The overlaied \emph{GOES} SXR 1--8 {\AA} flux variation reveals the M3.7 flare caused by the filament partial eruption.
(c)--(d): \emph{ASO-S}/FMG LOS magnetogram and filtergram of Fe {\sc i} 5324 {\AA} exhibiting the photospheric magnetic environment
of the erupting filament. The red contours outline the fragments of pre-eruption filament observed in H$\alpha$ channel.
(e)--(h): \emph{FY-3E}/EUVI 195 {\AA}, \emph{SATech-01}/SUTRI 465 {\AA}, \emph{SDO}/AIA 171 {\AA}, and 94 {\AA} images showing
the M3.7 flare in different channels around its peak time.
An animation (Figure1.mp4) covering 19:00 UT to 20:25 UT on December 30 is available online, which presents the filament
partial eruption in AIA 304, 171, 193, and 94 {\AA} channels. The animation's duration is 12 seconds.
}
\label{fig1}
\end{figure*}

The event of interest took place around the main sunspot of NOAA AR 13176 (see Figure \ref{fig1} and the associated online animation).
On 2022 December 30, this AR approached the solar disk center, and there were several filaments around the main sunspot of the AR.
As shown in Figures \ref{fig1}(a), (c), and (d), the filament studied here was located to the northwest of the sunspot, with its
south end rooted on the sunspot region with positive magnetic polarity and the north one on a negative-polarity region to the north
of the sunspot. Subsequently, this filament (hereafter ``pre-flare filament") erupted and produced an M3.7 flare (peaked at 19:38 UT)
with a smaller precursor C4.0 flare (peaked at 19:10 UT). Because the \emph{CHASE} observations (18:48--19:04 UT and 20:23--20:39 UT)
did not cover the main phase of the flare, we can only see the target region before and after the eruption with \emph{CHASE}/HIS
H$\alpha$ images. It is revealed that there was still a filament (hereafter ``post-flare filament") to be left in the flare core
region at 20:27 UT after the eruption of pre-flare filament (see Figure \ref{fig1}(b)). Figures \ref{fig1}(e)--(h) show that the
M3.7 flare produced by the filament partial eruption event exhibited typical compact flare loops and multiple flare ribbons observed
in different EUV channels with temperatures ranging from 0.5 MK to 6.3 MK. As a result, we can conclude that the pre-flare filament
underwent a partial eruption process and produced an M3.7 flare.

As introduced in Section \ref{sect1}, the partial eruption of solar filaments could be mainly produced through two scenarios:
(1) the vertical splitting of one MFR caused by internal magnetic reconnection, and (2) the eruption of upper MFR in a double-decker
system containing two separate vertically-distributed parts. It is obvious that the most significant difference between the two scenarios
is reflected in the magnetic configuration of the pre-eruption filament system. To unambiguously determine which mechanism is dominant
in the event reported here, three aspects of the pre-eruption filament are needed to consider at least: material flow pattern, magnetic
structure, and their temporal evolutions.

\subsection{Configuration of the double-decker filament system}\label{sect32}
\subsubsection{Material flow patterns in the double-decker filament system}
\begin{figure*}
\centering
\includegraphics [width=0.99\textwidth]{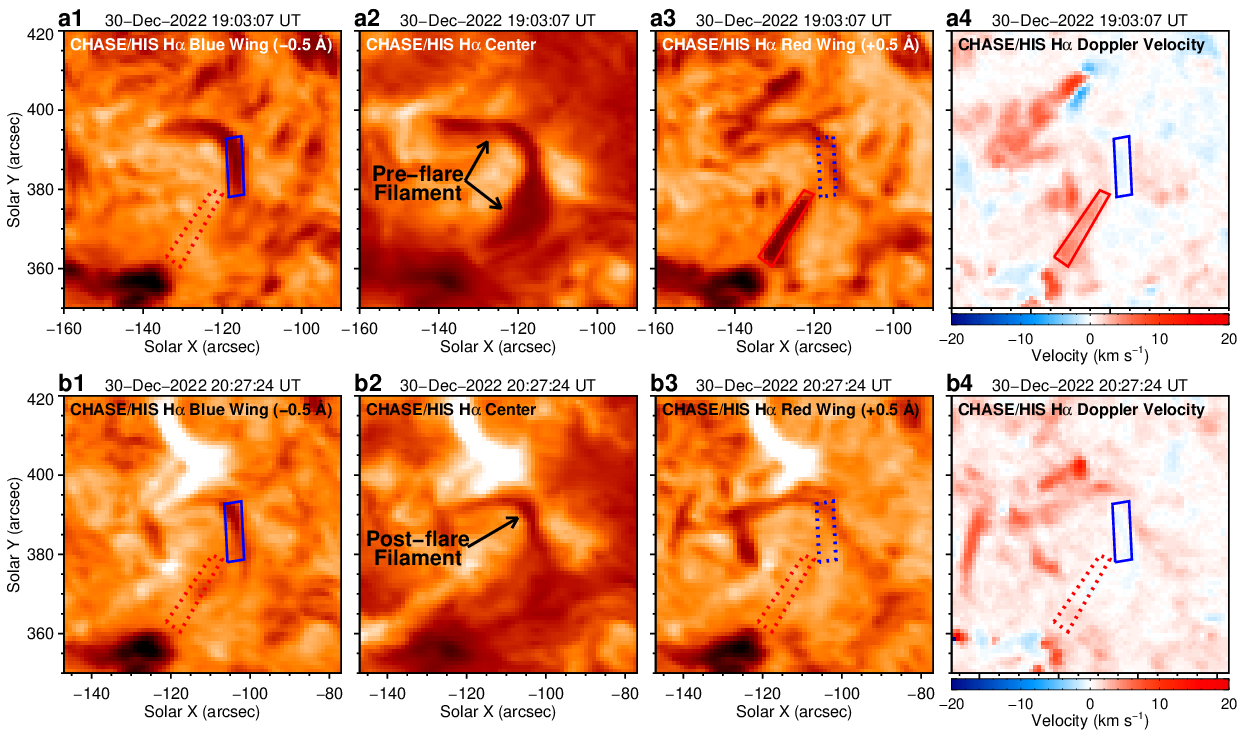}
\caption{Mass motions of the filaments before and after the partial eruption observed by the \emph{CHASE}/HIS in H$\alpha$ channel.
(a1)--(a4): Images of H$\alpha$ line center and two wings ($\Delta\lambda=\pm0.5$ {\AA}) and the corresponding Doppler velocity map
derived from H$\alpha$ spectra data exhibiting the material flows in filaments before the eruption. The blue and red quadrilaterals
mark the filament sections dominated by blueshift and redshift, respectively. The choice of solid and dotted lines depends on whether
the signal is clear.
(b1)--(b4): Similar to (a1)--(a4), but for the filament after the eruption.
}
\label{fig2}
\end{figure*}

To investigate mass motions of the filament system, we analyzed the \emph{CHASE}/HIS spectroscopic observations at the
H$\alpha$ waveband. As shown in Figures \ref{fig2}(a1)--(a3), before the eruption, the pre-flare filament
manifests as dark absorption features with different spatial distributions in the images of H$\alpha$ line center and two
wings. The filament system has a shape of the number ``7" in the H$\alpha$ center image and its south part is much wider
than the north part. In the H$\alpha$ blue wing (-0.5 {\AA}) image, only the north part of the filament is visible. While
in the H$\alpha$ red wing (+0.5 {\AA}) image, the southeast narrow part of the filament is obvious. Furthermore, the
Doppler velocity map derived from H$\alpha$ spectra data reveals that in the south part of the filament, strong redshift
signal (about 8 km s$^{-1}$) only appears in the east narrow part. The north region of the filament is partially
dominated by blueshift of 4--6 km s$^{-1}$ (see the blue and red quadrilaterals in panel (a4)).
Figures \ref{fig2}(b1)--(b4) show that after the partial eruption, the north part of the filament system and the
associated blueshift signal are still visible but with a much narrower spatial scale. Meanwhile, in the south region,
the filament absorption feature and redshift Doppler signal are all absent.

Material flows in a filament system can provide critical clues about its host magnetic structures. It is thus reasonable
to expect that filaments with different magnetic topologies will express different flow patterns. For example, in the
filament with a single uniform magnetic configuration, Doppler redshift and blueshift signals will be distributed randomly
due to counter-streaming flows in different filament threads \citep{1998Natur.396..440Z,2020NatAs...4..994Z} or regularly
detected at the opposite sides of the filament spine over a large spatial range due to the rotation motion of the filament
material along the twisted field of an MFR \citep{2019ApJ...872..109A}. But in the event reported here, the structured
distributions of redshift and blueshift within the pre-flare filament do not show typical characteristics of a single uniform
magnetic configuration as mentioned above. Instead, the fact that redshift only appears in the east narrow part of the south
region indicates that there could be two separate material flow patterns within the filament system, at least in the south
region. Furthermore, the disappearance of the southeast narrow part of the filament system dominated by redshift and the
invariableness of the north part dominated by blueshift after the partial eruption suggest that the region of redshift and
the region of blueshift could be two separate material flow systems within the filament.

\begin{figure*}
\centering
\includegraphics [width=0.99\textwidth]{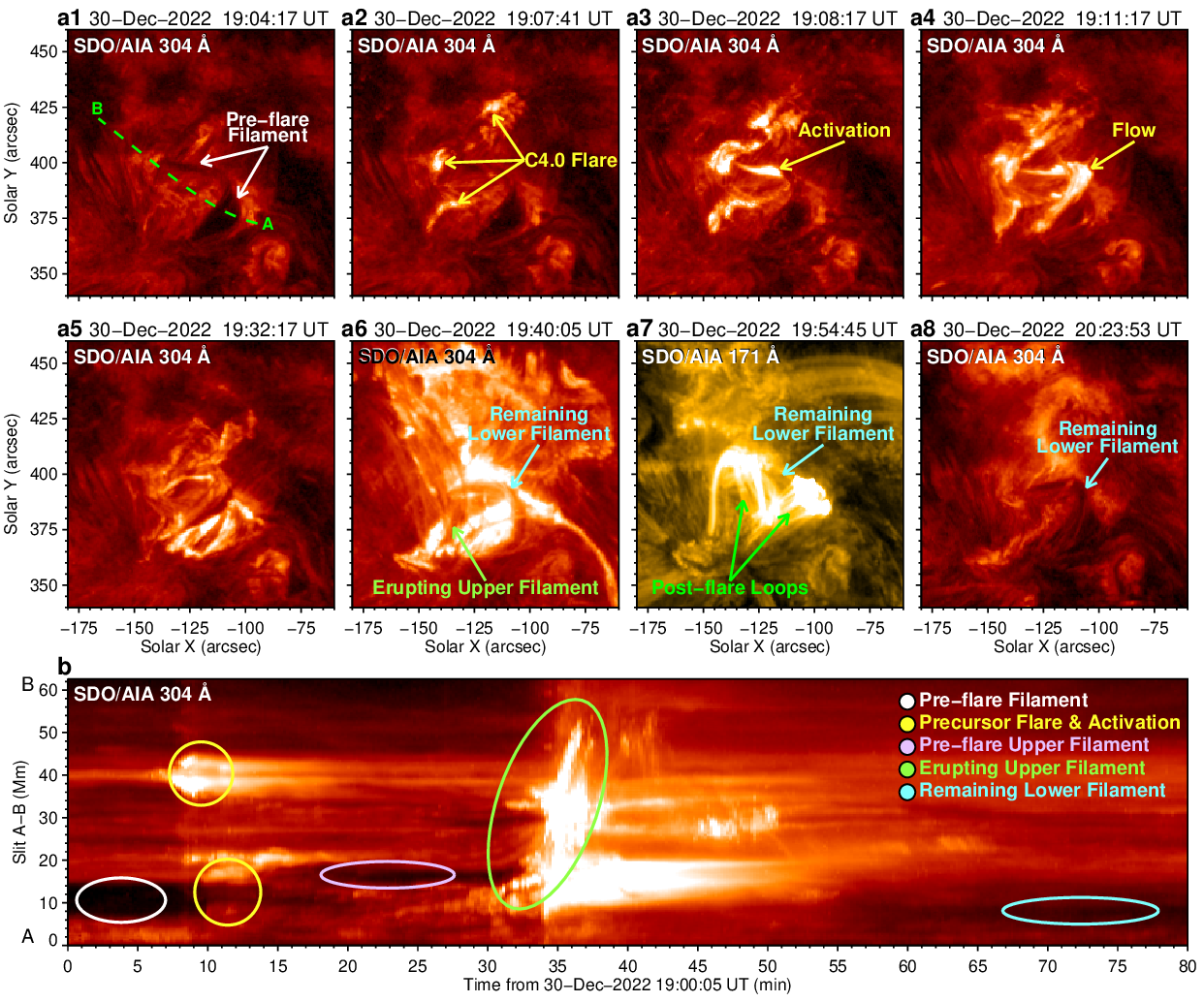}
\caption{Temporal evolution of the double-decker filament system before and during its partial eruption.
(a1)--(a8): Sequence of \emph{SDO}/AIA 304 and 171 {\AA} images showing the partial eruption process of the double-decker
filaments and the resultant post-flare loops. The slit ``A--B" in (a1) approximates the moving direction of erupting
filament section.
(b): Time-distance map derived from the 304 {\AA} images along the slit ``A--B" shown in (a1).
}
\label{fig3}
\end{figure*}

Based on the successive EUV images obtained by \emph{SDO}/AIA, we further investigated the material structure of the
filament system by analyzing its temporal evolution during the M3.7 flare. In AIA 304 {\AA} observations, the pre-flare
filament shows the similar morphology as that in the images of H$\alpha$ line center (see Figure \ref{fig3}(a1)).
Around 19:07 UT, a C4.0 flare occurred to the northeast of the filament system and then evolved into three
separate compact bright patches (panel (a2)). Meanwhile, another weak brightening appeared at the site on the west
of the filament's northern footpoint and then illuminated the whole north part of the filament system around 19:08 UT
(panel (a3)). The brightening then propagated laterally along the filament spine and flipped from bottom to top after
rounding the turning point of the shape of ``7", forming a southeastward material flow. This flow illuminated the
southeast narrow region of the filament and the structured distribution of redshift (panel (a4)). About 20 minutes
later, a compact brightening was detected beneath the south part of the filament and then the whole filament was
activated (panel (a5)), resulting in the final eruption of the upper part of the filament system (panel (a6)).
In subsequent AIA 171 and 304 {\AA} observations, we can see that this eruption produced a set of post-flare loops
outlining the initial ``7" shape of the erupting filament. Moreover, under the post-flare loops, there was still a
filament remaining where the erupting one was located (panels (a7) and (a8)). We speculate that it is the material
flow in the remaining filament section that produces the region dominated by blueshift in Figure \ref{fig2}.

Along the moving direction of erupting filament section, we also constructed a time-distance map through AIA 304 {\AA} images
as shown in Figure \ref{fig3}(b). It is obvious that the south region of the pre-flare filament system was relatively wide in
the beginning, which was then activated by surrounding small-scale flare. As a result, the east part (the upper part) of
this region was lifted a little while the the west part (the lower part) was intermittently but successively heated, which
thus did not show dark absorption features. Subsequently, the upper filament section erupted northeastwards with a typical
two-phase evolution: slow rise followed by a rapid eruption. About half an hour after the eruption, the west part (the lower
part) of the filament system cooled down and showed dark absorption features again. Combining the Doppler velocity distribution
and temporal evolution of the filament, we propose that there were two independent material flow systems within the pre-flare
filament and the upper part eventually erupted while the lower part remained, which might be supported by two up-and-down MFRs,
i.e., double-decker MFR system \citep{2012ApJ...756...59L}.

\subsubsection{Magnetic topology of the double-decker filament system}
\begin{figure*}
\centering
\includegraphics [width=0.99\textwidth]{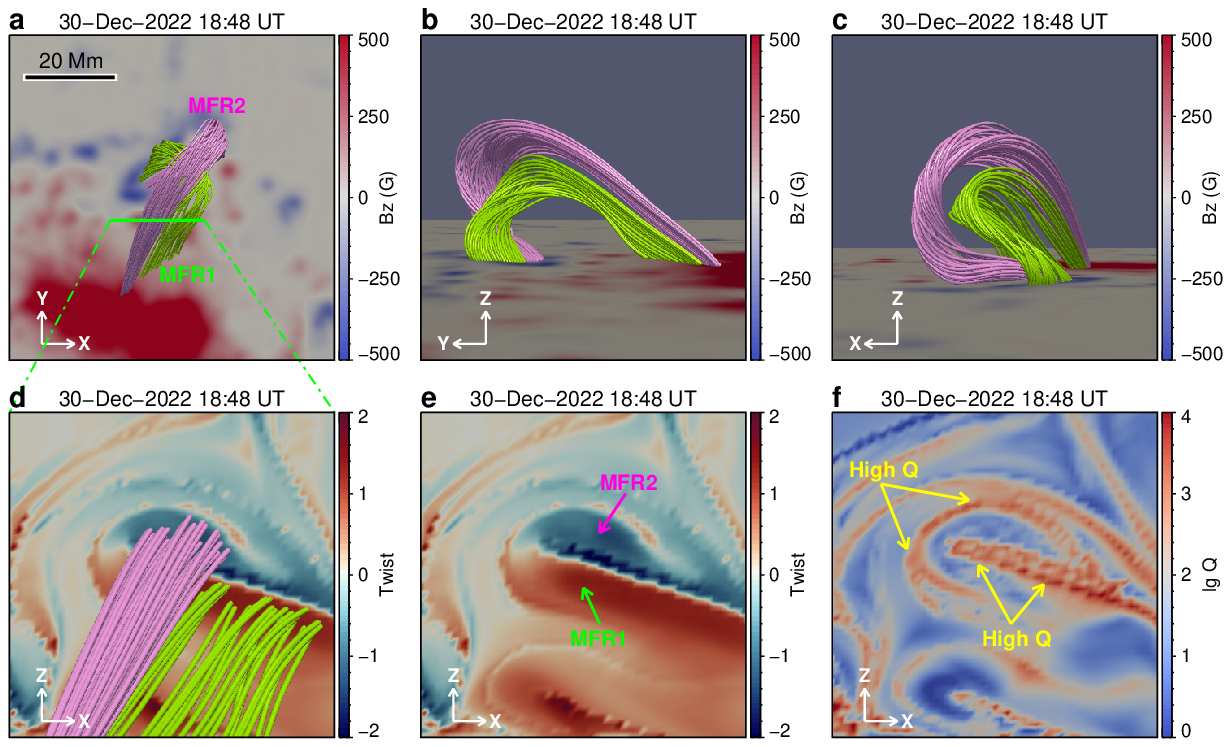}
\caption{Three-dimensional (3D) magnetic topology of the filaments before the partial eruption revealed by NLFFF extrapolation.
(a)--(c): Top and side views of two MFRs (MFR1 and MFR2) composing a double-decker configuration.
(d)--(f): Distributions of magnetic twist $T_{w}$ and logarithmic $Q$ in the vertical plane based on the green cut denoted in (a).
}
\label{fig4}
\end{figure*}

To verify the speculation about the double-decker MFR structure, we further extrapolated 3D magnetic topology of the filament system
through NLFFF method based on the photospheric vector magnetic fields of \emph{SDO}/HMI at 18:48 UT on 2022 December 30, just before
the occurrence of the M3.7 flare. Moreover, we calculated the twist number $T_{w}$ and the squashing factor $Q$ of the reconstructed
fields. Then we can obtain the $T_{w}$ and $Q$ distribution maps in the photospheric or selected vertical planes. According to the
photospheric twist map, we plotted the magnetic field lines across the photosphere where $\mid$$T_{w}$$\mid$ $\geq$ 1.0 around the
footpoint of the filament system. As shown in Figure \ref{fig4}, there are indeed two MFRs in the flare core region: MFR2 with
$T_{w}<$--1.0 above MFR1 with $T_{w}>$1.0, forming a double-decker MFR configuration \citep{2012ApJ...756...59L}. In the south region
of this configuration, higher MFR2 is located to the east of MFR1 in the plane of sky (POS). But in the north region, MFR2 passes
MFR1 from below and roots in a negative-polarity region located to the west-south of the footpoint of MFR1.

\begin{figure*}
\centering
\includegraphics [width=0.95\textwidth]{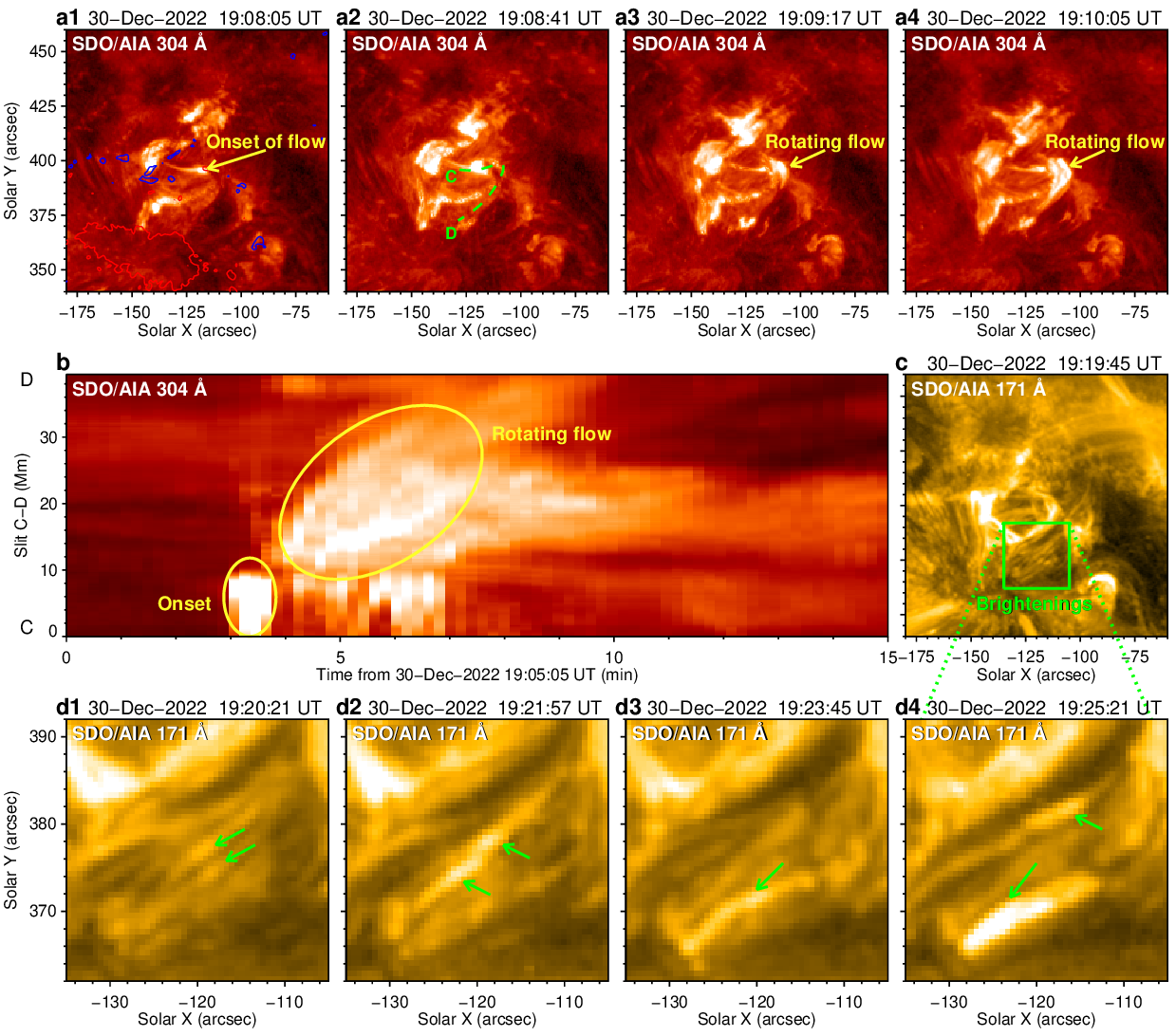}
\caption{Material flow and transient brightenings within the double-decker filament system before its partial eruption.
(a1)--(a4): Sequence of \emph{SDO}/AIA 304 {\AA} images showing the material flow rotating around the lower filament (MFR1). The slit
``C--D" in (a2) approximates the moving direction of the flow.
(b): Time-distance map derived from the 304 {\AA} images along the slit ``C--D" shown in (a2).
(c): 304 {\AA} image showing the filament system after being disturbed by the flow.
(d1)--(d4): Sequence of \emph{SDO}/AIA 171 {\AA} images showing the frequent brightenings in the south region of the filament system
before its eventual eruption.
}
\label{fig5}
\end{figure*}

In Figures \ref{fig5}(a)--(b), the heated material flow rotating the filament system is shown clearly. We can see that the
flow originated beneath the north part of the filament, then flipped from bottom to top, rotating around the lower filament, and
eventually deposited at a site on the east of the filament's southern footpoint. Combining the EUV observations and NLFFF results
shown in Figures \ref{fig3}--\ref{fig5}, we can conclude that they are consistent with each other according to the following two facts:
1) The extrapolated magnetic structure and the observed filament material flow share similar footpoint positions; 2) MFR2 rotates
around MFR1 in their north part in a similar way that the bright flow in the upper filament rotates around the lower filament. As
a result, we suggest that MFR2 represents the magnetic structure of the erupting upper filament and MFR1 corresponds to the remaining
lower filament. About the consistency between the observations and NLFFF results, we would like to note that the field lines shown
here are the magnetic field lines with absolute twist number larger than 1.0, which means that there are still other less twisted
field lines of the filament that are not plotted. Therefore, the magnetic structure shown in Figure 4 might not exactly match the
observed filament system in space.

Along the green cut marked in Figure \ref{fig4}(a), we make $T_{w}$ distribution maps in the vertical (x-z) plane and show it in
panels (d) and (e). One can see that there are two regions with high positive and negative $T_{w}$ in this vertical twist map,
which correspond to the cross sections of MFR1 and MFR2, respectively. High $Q$ ribbons are located along the edge of the two
high twist regions and the interface between them as shown in panel (f). Although the high-$Q$ regions are usually locations
favorable for magnetic reconnection, the two MFRs of the double-decker structure reported here have opposite magnetic twist, which
means that the field line of the two MFRs at the intersecting boundary between them would have the same direction. As discussed by
\citet{2021ApJ...909...32P}, such special double-decker topology could be more stable than the magnetic configurations having the
same sign of twist originally proposed by \citet{2012ApJ...756...59L} because the two vertically-distributed parts are not separated
by a hyperbolic flux tube (HFT) where a current layer likely develops. As a result, it is reasonable that the lower filament of this
double-decker configuration apparently did not undergo any significant morphological change during the partial eruption. However,
during the 20 minutes before the onset of the partial eruption, we noticed that there were frequent local EUV brightenings appearing
around the south region of the filament system (Figures \ref{fig5}(c)--(d)), which might be due to the interaction within
the two vertically-distributed filaments (or MFRs). Although the precursor event driving the filament partial eruption is an
interesting topic, it is beyond the scope of the present work and will be discussed in detail in our future study.

Since the two MFRs composing a double-decker configuration have similar footpoints, it is reasonable to believe that they have the
same sign of helicity and are formed successively through similar way, e.g., direct emergence from below the photosphere
\citep{2001ApJ...554L.111F,2008ApJ...673L.215O}, magnetic reconnection of sheared arcades in the corona \citep{1989ApJ...343..971V,
2003ApJ...585.1073A,2014ApJ...789...93C}, or through splitting process of a single MFR \citep{2001ApJ...549.1221G,2014ApJ...792..107K,
2021ApJ...909...32P}. However, the double-decker configuration reported here consists of two MFRs with opposite magnetic twist,
which is not consistent with the magnetic configurations proposed by \citet{2012ApJ...756...59L}, where the MFRs or sheared arcades
have the same sign of twist. Then, a question naturally arises as to how such double-decker configuration containing two MFRs with
opposite twist in equilibrium can form. In the next subsection, we will explore the possible formation of such configuration
and try to find observational evidence of the two vertically-distributed filaments of being opposite twist.

\subsection{Formation of the double-decker filament system}\label{sect33}
\begin{figure*}
\centering
\includegraphics [width=0.99\textwidth]{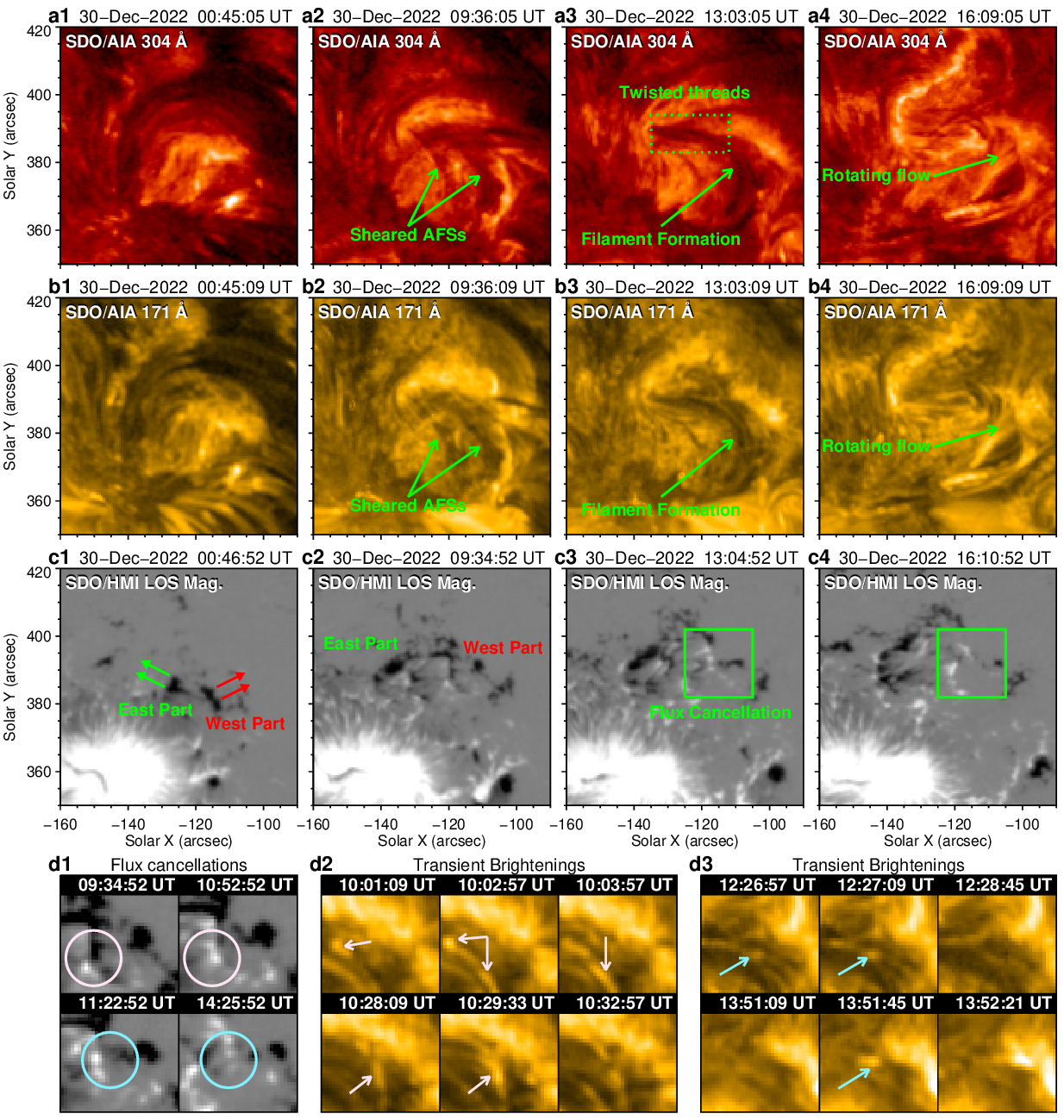}
\caption{Formation of the double-decker filament system.
(a1)--(b4): Sequences of AIA 304 and 171 {\AA} images showing the formation of the double-decker filament system.
(c1)--(c4): Sequence of HMI LOS magnetograms exhibiting the opposite shearing motions of the northern footpoints
of the initial filament structures and flux cancellation during the double-decker filament formation. The green and
red solid arrows in (c1) denote the moving directions of the east and west parts of the northern footpoints. The green
squares mark the region where magnetic flux cancellation happens successively, as well as the field view of
panels (d1)--(d3).
(d1)--(d3): Sequences of HMI LOS magnetograms and AIA 171 {\AA} images exhibiting two flux cancellation processes
and associated transient brightenings around the west part of the northern footpoints of the initial filament structures
(i.e., the northern footpoint of the initial structure of upper MFR2). The circles and arrows denote the sites of
the flux cancellation and brightenings.
An animation (Figure6.mp4) of AIA 171, 304 images and HMI LOS magnetograms, covering 00:00 UT to 19:00 UT on
December 30, is available in the online journal to present the formation of the double-decker filament system.
The animation's duration is 11 seconds.
}
\label{fig6}
\end{figure*}

To investigate the formation of the double-decker configuration containing two MFRs with opposite twist, we firstly
analyzed the pre-eruption \emph{SDO}/AIA 304, 171 {\AA} observations and \emph{SDO}/HMI LOS magnetograms on 2022 December
30 (see Figure \ref{fig6} and the corresponding animation). One can see that at the beginning of December 30, there were
only several north-south arch filament systems (AFSs) in the region of interest, connecting the sunspot region with positive
magnetic polarity and a negative-polarity region to the north of the sunspot (panel (a1)). Then, the east part of
north end of these AFSs kept moving northeastwards and the AFSs thus became more and more sheared, which eventually evolved
into a filament with a shape of ``7" around 13:03 UT (panels (a2)--(a3)). During the following several hours,
this filament gradually grew up to have the shape of the pre-flare filament as shown in Figures \ref{fig2} and \ref{fig3}.
As shown in Figure \ref{fig6}(a4), around 16:09 UT, a bright material flow appeared to rotate around the filament, similar
to the flow rotating the lower filament occurring before the onset of the filament partial eruption. It indicates that the
double-decker configuration had probably completely formed at that time and the flow was along the upper MFR2.

Meanwhile, HMI LOS magnetograms reveal that the negative-polarity region to the north of the sunspot can roughly be divided
into two patches: east part and west part (Figure \ref{fig6}(c1)). The east part is the footpoint of the north leg of the
filament and kept moving northeastwards while the west part kept moving northwestwards (see the green and red arrows).
Small-scale dipolar magnetic fields with direction of west-east successively emerged between the two parts. As a result, the
positive patches of the emerging dipolar fields kept cancelling with the existing west part negative fields (see Figures
\ref{fig6}(c2)--(c4)). Meanwhile, transient brightenings around sites of the flux cancellation were intermittently
observed in AIA EUV images (see Figures \ref{fig6}(d1)--(d3)). It is widely accepted that the shearing motion of AFSs
observed here would result in the formation of filament through magnetic reconnection between sheared arcades. But such
process can not produce a double-decker configuration of two MFRs with opposite magnetic twist as reported here. We speculate
that the formation of such double-decker filament system could be caused by another process related to the magnetic flux
cancellation and brightenings around the west part negative fields shown in Figures \ref{fig6}(c2)--(d3).

\begin{figure*}
\centering
\includegraphics [width=0.99\textwidth]{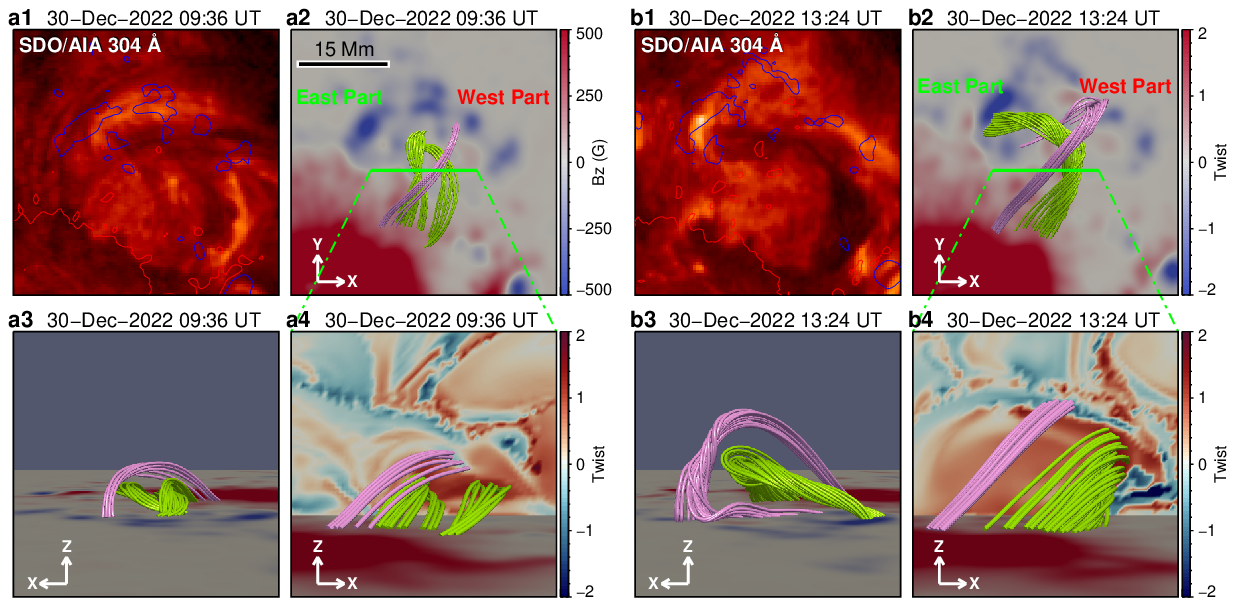}
\caption{Temporal evolution of magnetic and emission structures during the formation of the double-decker filament.
(a1)--(a4): AIA 304 {\AA} image and NLFFF extrapolation results at 09:36 UT exhibiting the filament system and its
3D magnetic topology at the early stage of formation. The green and pink field lines represent initial magnetic
topology of lower MFR1 and upper MFR2, respectively.
(b1)--(b4): Similar to (a1)--(a4), but for the time point of 13:24 UT at the late stage of formation.
}
\label{fig7}
\end{figure*}

\begin{figure*}
\centering
\includegraphics [width=0.9\textwidth]{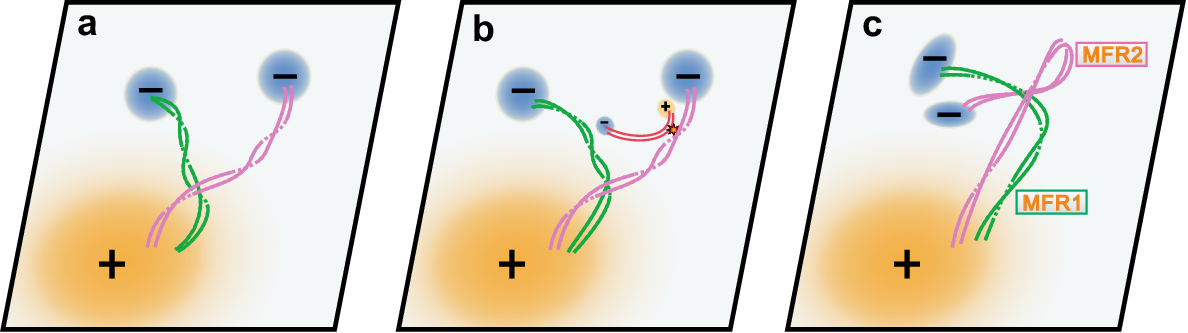}
\caption{Schematic diagram depicting the formation of double-decker configuration consisting of two MFRs with opposite magnetic
twist. The green and pink twisted curves in (a) represent initial magnetic topologies of the two MFRs: lower MFR with direction of
north-east to south-west and positive magnetic twist; upper MFR with direction of north-west to south-east and negative magnetic
twist. The red curves in (b) delineate emerging dipolar magnetic fields with direction of west-east between the north footpoints
of the two MFRs. The star symbol marks the site of magnetic reconnection between the emerging dipolar magnetic fields and the
north-west leg of the upper MFR, producing eventually the double-decker configurations consisting of MFR1 and MFR2 with similar
footpoints but opposite magnetic twist (see twisted curves in (c)).
}
\label{fig8}
\end{figure*}

To verify the speculation about the formation of the double-decker structure, we further investigate temporal evolution
of 3D magnetic structures of the filament system based on NLFFF extrapolations with a time sequence. Figures \ref{fig7}(a1)--(a4)
show that at 09:36 UT on December 30, two sets of twisted magnetic structures with direction of southwest to northeast and
positive $T_{w}$ were extrapolated (see the green lines), corresponding to the two sets of sheared AFSs in the target region.
Meanwhile, we should note that there was another sheared structure with a different direction of southeast to northwest and
negative $T_{w}$ (see the pink lines) above the AFS structures. And the northwest footpoint of this upper structure was
located at the west negative-polarity patch as shown in Figures \ref{fig6}(c1)--(c2). From Figures \ref{fig7}(b1)--(b4),
one can see that, when the sheared AFSs evolved into a ``7"-shaped filament at 13:24 UT, a MFR structure with the similar
shape and highly positive $T_{w}$ also formed. As for the magnetic structure with negative $T_{w}$ at a higher altitude,
it became more twisted and its northwest footpoint partially moved to the east negative-polarity region near the north end of
the lower MFR. In the meantime, magnetic flux cancellation and the associated EUV brightenings intermittently occurred around
the northwest end of this upper twisted structure as shown in Figures \ref{fig6}(d1)--(d3).

Temporal evolution of the filament system shown in Figures \ref{fig6} and \ref{fig7} provides critical clues for the formation
of the double-decker configuration of two MFRs with opposite magnetic twist. We infer that the two MFRs of the double-decker
configuration originated from two magnetic systems with different initial topologies: i.e., different directions, connections,
and opposite magnetic twist. But subsequent interactions with the emerging fields changed the magnetic connection of initial
structure of the upper MFR2 by moving its north end eastwards to a region near the north end of the lower MFR1. The continuous
flux cancellation and associated EUV brightenings around the northwest end of the initial structure of the upper MFR2
(Figures \ref{fig6}(d1)--(d3)) provide a solid evidence for the existence of magnetic reconnection driving the
magnetic connection change of MFR2. In this scenario, it is acceptable that MFR1 and MFR2 have opposite magnetic twist when
they eventually evolved into a double-decker MFR configuration with similar footpoints. To illustrate more intuitively this
physical scenario for the formation of double-decker filament system, we draw a schematic diagram as shown in Figure \ref{fig8}.

It is worth noting that for the large-scale filaments observed by high-resolution ground-based telescopes, one can clearly
determine the sign of their magnetic twist according to the filament threads or heated plasma flows along the twisted field
lines of the filament. But in the event reported here, the double-decker filaments of interest are small-scale active region
filaments and show no obvious signals of twisted threads or flows in the \emph{CHASE} or \emph{SDO} observations. Moreover,
because the lower filament was located below the upper one and they were near the solar disk center, it is difficult to
determine whether the two parts have opposite twist through the observations from a top-down perspective. However, during the
formation process of the double-decker configuration, we find some observational evidence of the two vertically-distributed
filaments of being opposite twist as follows: 1) As shown in Figure \ref{fig6} and the corresponding animation, directions of
shearing motion forming the initial structures of lower and upper filaments/MFRs are opposite, which would naturally result in
the opposite twist of the two filaments/MFRs. 2) When the upper filament was not formed yet, we find twisted threads of the
lower filament with a positive sign of magnetic twist at 13:03:05 UT (Figure \ref{fig6}(a3)), which is consistent with the
sign of magnetic twist of the lower MFR revealed by the NLFFF modeling.

Regarding the formation of the double-decker configuration containing two MFRs with the same magnetic twist in equilibrium,
\citet{2012ApJ...756...59L} proposed two possible scenarios: (1) After the formation of upper MFR above the PIL, the lower branch
emerges from below the photosphere at the same site (hereafter ``emerging model"). (2) Both branches originally belong to a
single MFR or flux bundle and are then separated into two parts (hereafter ``splitting model"). The emerging model can be
supported the observations that an MFR can directly emerge under a pre-existing filament \citep{2008ApJ...673L.215O}. As for
the splitting model, it is motivated by the ``partial eruption" scenario proposed by \citet{2001ApJ...549.1221G}, in which
internal magnetic reconnection within an MFR can split it into two MFRs with the same handedness, and further verified by
following observations and numerical modeling results \citep{2014ApJ...792..107K,2018NewA...65....7T,2021ApJ...909...32P}.
The formation of the initial MFR in the two scenarios can be attributed to the reconnection between two groups of sheared
arcades near the PIL driven by shearing, converging, and rotation motions of their magnetic footpoints
\citep{2014ApJ...789...93C,2018ApJ...856...79Y,2018A&A...619A.100H,2022ApJ...933..200Z}. In the present work, we propose
another scenario to explain the formation of the double-decker MFRs with opposite magnetic twist reported here. As shown
by Figure \ref{fig8}, in this scenario, magnetic reconnection between the newly emerging fields and initial structure of the
upper MFR plays a key role in the formation of eventual double-decker configuration by moving the footpoint of the involved
MFR. Such footpoint evolution driven by magnetic reconnection between solar filaments and their surrounding structures has been
frequently reported \citep{2016NatPh..12..847L,2017ApJ...838..131Y,2018ApJ...853L..26H,2021ApJ...920...77G,2023arXiv230909414S}.
Moreover, we suggest that such scenario might also apply to the formation of typical double-decker configuration with the same
magnetic twist.

\section{Summary}\label{sect4}
Based on the high-resolution observations from the \emph{CHASE}, \emph{SDO}, and \emph{ASO-S}, we investigate a typical filament
partial eruption event, present integrated evidence for the double-decker MFR configuration of pre-eruption filament, and propose
a new scenario for its formation. The main results are summarized as follows:
\begin{enumerate}
\item On 2022 December 30, a filament was located to the northwest of the main sunspot of NOAA AR 13176 and its subsequent partial
eruption produced an M3.7 flare. The \emph{CHASE} H$\alpha$ spectroscopic observations reveal distinct structured distribution of
Doppler velocity within the pre-flare filament, where redshift only appeared in the east narrow part of the south filament region
and then disappeared after the partial eruption while the north part dominated by blueshift remained. Combining the Doppler
velocity distribution and temporal evolution of the filament, we infer that there were two independent material flow systems
within the pre-flare filament, which might be supported by a double-decker configuration.

\item NLFFF extrapolations reveal that there are indeed two vertically-distributed MFRs in the flare core region: MFR2 with
$T_{w}<$--1.0 above MFR1 with $T_{w}>$1.0, forming a double-decker MFR configuration. In the south region of this
configuration, higher MFR2 is located to the east of MFR1 in the POS. But in the north region, MFR2 passes MFR1 from below
and roots around the footpoint of MFR1. This magnetic configuration is consistent with the observations mentioned above, and we
suggest that MFR2 represents the magnetic structure of the erupting upper filament and MFR1 corresponds to the remaining lower
filament.

\item The double-decker configuration reported here consists of two MFRs with similar footpoints but opposite magnetic twist.
To explain the formation of such type of double-decker MFR configuration, we proposed a new scenario as follows: the two MFRs
of this double-decker configuration originate from two magnetic systems with different initial connections and opposite magnetic
twist. But subsequent magnetic reconnection between the initial structure of the upper MFR and surrounding newly-emerging fields
gradually results in the motion of the footpoint of the upper MFR to a region around the footpoint of the lower MFR, thus leading
to the formation of eventual double-decker configuration.
\end{enumerate}

It is worth noting that although the NLFFF extrapolation results and observations of HMI LOS photospheric magnetograms
and AIA EUV images support our scenario for the formation of the double-decker MFR configuration reported here, the initial
structure of overlying MFR2 is difficult to identify clearly in the AIA observations. The absence of the precursor structure
of MFR2 could be caused by its low plasma density at the early stage. After interacting with the lower emerging field filled
with plenty of cold plasmas, the newly-formed MFR2 will be injected by a large quantity of heated plasma and then cooled down,
forming the upper section of the double-decker filament system distinct in the EUV and H$\alpha$ imaging observations.
In addition, there is another important topic that is not discussed in this paper: how was such double-decker filament system
activated and then evolved into partial eruption, producing an M3.7 flare? We have noticed that in AIA EUV observations, there
were frequent local brightenings and bright threads appearing in the south region of filament during the 20 minutes before its
eventual eruption. The evolution of the double-decker filament system and driving mechanism of its partial eruption will
be discussed in detail in our future paper (Paper II, in preparation).

\acknowledgments
The authors are cordially grateful to the anonymous referee for the constructive comments and suggestions.
The data used here are courtesy of the \emph{CHASE}, \emph{ASO-S}, \emph{SDO}, \emph{GOES}, \emph{SUTRI}, and \emph{FY-3E} science teams.
The authors are supported by the National Key R\&D Program of China (2022YFF0503800, 2022YFF0503000, and 2019YFA0405000), the Strategic
Priority Research Program of CAS (XDB0560000 and XDB41000000), the National Natural Science Foundation of China (12273060, 12222306, 12333009,
12127901, 12073042, 12233012, 11921003, and 11903050), ``Integration of Space and Ground-based Instruments" project of CNSA, the Youth
Innovation Promotion Association CAS (2023063), the Open Research Program of Yunnan Key Laboratory of Solar Physics and Space Science (YNSPCC202211), BJAST Budding Talent Program (23CE-BGS-07), and Yunnan Academician Workstation of Wang Jingxiu (No. 202005AF150025).
\emph{CHASE} mission is supported by CNSA. \emph{SDO} is a mission of NASA's Living With a Star Program. \emph{ASO-S} mission is supported
by the Strategic Priority Research Program on Space Science of CAS, Grant No. XDA15320000. \emph{SUTRI} is a collaborative project conducted
by the National Astronomical Observatories of CAS, Peking University, Tongji University, Xi'an Institute of Optics and Precision Mechanics
of CAS and the Innovation Academy for Microsatellites of CAS.



\bibliography{ref}{}
\bibliographystyle{aasjournal}

\end{document}